\documentclass[12pt,preprint]{aastex}

\slugcomment{Submitted to the AJ}

\shorttitle{RR Lyrae stars in the Sgr Tidal Stream}
\shortauthors{Vivas, Zinn \& Gallart}

\begin{document}

\title{VLT Spectroscopy of RR Lyrae Stars in the Sagittarius Tidal 
Stream}

\author{A. Katherina Vivas}
\affil{Centro de Investigaciones de Astronom{\'\i}a (CIDA), Apartado Postal
264, M\'erida 5101-A, Venezuela}
\email{akvivas@cida.ve}

\author{Robert Zinn}
\affil{Department of Astronomy, Yale University, PO Box 208101, New Haven,
CT 06520-8101, USA}
\email{zinn@astro.yale.edu}

\and

\author{Carme Gallart}
\affil{Instituto de Astrof{\'\i}sica de Canarias (IAC), Calle V{\'\i}a 
L\'actea, E-38200, La Laguna, Tenerife, Spain}
\email{carme@ll.iac.es}

\begin{abstract}

Sixteen RR Lyrae variables from the QUEST survey that lie in the
leading arm of the tidal stream from the Sagittarius dSph galaxy have
been observed spectroscopically to measure their radial velocities and
metal abundances.  The systemic velocities of 14 stars, which were
determined by fitting a standard velocity curve to the individual
measurements, have a sharply peaked distribution with a mean of
$33$ km/s and a standard deviation of only 25 km/s.  The [Fe/H]
distribution of these stars has a mean of $-1.76$ and a standard
deviation of 0.22.  These measurements are in good agreement with
previous ones from smaller samples of stars.  The mean metallicity is
consistent with the age-metallicity relation that is observed in the
main body of the Sgr dSph galaxy.

The radial velocities and the distances from the Sun of these stars
are compared with recent numerical simulations of the Sgr streams that
assume different shapes for the dark matter halo.  Models that assume
a oblate halo do not fit the data as well as ones that assume a
spherical or a prolate distribution.  However, none of the fits are
completely satisfactory.  Every model fails to reproduce the long
extent of the stream in right ascension ($36\degr$) that is seen in
the region covered by the QUEST survey.  Further modeling is required
to see if this and the other mismatches between theory and observation
can be removed by judicial choices for the model parameters or instead
rule out a class of models.
\end{abstract}


\section{INTRODUCTION}

From the time of its discovery \citep{iba94}
the Sagittarius (Sgr) dwarf spheroidal
(dSph) galaxy has been considered the prototypical example of the
tidal destruction and assimilation of a small galaxy into the halo of
a larger one.  Long tidal streams of stars from the Sgr dSph, which
were predicted by theoretical models of the destruction process 
\citep[e.g.][]{joh99}, have now been observed in several
directions with different stellar tracers by the recent large-scale
surveys of the halo.  An all-sky view is provided by the M giants
detected by the 2MASS survey, which reveals two prominent tidal
streams, one leading and one trailing the main body of the dSph galaxy
\citep{maj03}. In addition to these metal-rich stars, the streams
also contain numerous RR Lyrae stars \citep{ive00,viv01,vz04}, which,
as we show later, are metal poor in addition to being very old.

The detailed study of the Sgr tidal streams is important for several
reasons.  First, it gives us a close-up look into the process of tidal
destruction and galaxy merging, which may be responsible for the
formation of the halos of disk galaxies, as suggested by the
hierarchical models of galaxy formation and by numerous studies of the
Milky Way's halo over the past 25 years \citep[see review by][]{fre02}.  
Second, comparisons between
the stellar populations at different points in the streams and the
main body of Sgr may reveal interesting facets of galaxy evolution in
the presence of strong tidal forces, in particular if dynamical models
can provide the chronology of the formation of the streams.  Finally,
tidal streams are powerful probes of the shape of the potential of the
Milky Way \citep{joh98}, which at the large galactocentric distances
of the Sgr streams may be dominated by the dark matter halo.  There has
been some recent debate, however, over the usefulness of the Sgr streams
for this purpose. N-body simulations of modern hierarchical galaxy
formation picture predict that dark matter halos should be flattened
by appreciable amounts \citep[e.g.][]{bul02}. The
fact that the Sgr tidal streams can be traced with both Carbon stars
and M giants along a narrow great circle on the sky has been used as
an argument in favor of an almost spherical halo \citep{iba01,maj03};
otherwise the stars in the stream would disperse and the resulting
stream would be much wider and perhaps not even recognizable as a
feature above the background of halo stars \citep{may02}.  While this
alone may not rule out the hierarchical models, the apparent
deviation of the most familiar dark matter halo from the predicted
norm is somewhat discomforting.  More recently \citet{hel04} has
argued that only old streams will be dispersed by the precession
generated by a flattened potential and that the Sgr streams are too
young dynamically, with most of the stars stripped out in the past few
perigalacticon passages \citep{hel01,hel04,mar04}, for this to have
occurred.  The theoretical models of the streams under different
assumptions of halo shape, \citep[e.g.][]{hel04}, are in fact different
on small scales.  The precisions that are necessary to test
observationally these models are achievable with RR Lyrae variables,
but with few other halo tracers.

The spatial distribution of the Sgr streams is well documented by the
investigations of
\citet{yan00,ive00,iba01,viv01,mar01,new02,maj03,mar04} and
\citet{vz04}. The tidal streams should be also coherent
structures in velocity space as well. The most complete work on the subject to
the date is that of \citet{maj04}, who obtained radial velocities for
$\sim 300$ M giant candidates, most of them belonging to the Sgr
trailing (Southern Galactic Hemisphere) stream.  They find that the feature is
indeed a coherent structure in velocity space.

In this paper, we present a spectroscopic study of RR Lyrae stars
from the QUEST survey \citep{viv04} belonging to the Sgr leading arm
(Northern Galactic Hemisphere). These observations provide a unique
set of detailed observations in this part of the stream: location on
the sky, distance from the Sun, spatial dimensions, radial velocities
and metallicities. Many of these parameters are better determined in
this region than in any other part of the stream. This is because RR
Lyrae stars are excellent standard candles, which enables us to
determine their distances with precisions of $\sim 6\%$.  The
precisions obtained with other stellar tracers are substantially
worse, for example they are $20-25\%$ for K and M giants \citep{doh01,maj03}.
It is important to emphasize that our sample is made of {\it
bona-fide} RR Lyrae stars (not just candidates), which were discovered
using well-sampled light curves.  The purpose of our observations was
to constrain models of the disruption of the Sgr galaxy, and we show
that none of recent models of Sgr streams are consistent with these
observations.

This paper is organized as follows: in \S2 we describe the target
selection, spectroscopic observations and data reduction. Section \S3
describes the method for obtaining radial velocities of these
pulsating stars, and \S4 explains the determination of metal
abundances.  In \S5, we compare our results with different theoretical
models of the Sgr streams.  We briefly summarize our results in \S6.

\section{THE DATA}

\subsection{Target Selection}

The distribution on the sky of the 85 RR Lyrae variables in the QUEST
survey \citep{viv04} that are probably part of the Sgr stream covers
about $36\degr$ in right ascension, from $13\fh 0$ to $15\fh 4$
\citep{viv01,vz04}.  To confirm the size of the stream and
to detect any gradients in radial velocity with position, we selected
for spectroscopic observation stars from this sample that lie along
the whole length of stream and that span its range in magnitude (see
Figure~\ref{fig-targets}).  The 16 stars that were observed have a
mean magnitude of $V=19.27$ and range from 18.88 to 19.66.  These
stars lie approximately 50 kpc from the Sun.  The light curves of
these stars, which with their periods classify them as type
ab\footnote{The QUEST survey is not able to measure reliably the
smaller amplitude type c variables at these faint magnitudes
\citep{viv04}}, are well defined by the QUEST observations.  This
suggests that the Blazhko effect is not present.  The
ephemerides from the QUEST observations should be adequate for
predicting the phases of the spectroscopic observations.

\subsection{Observations}

During the pulsation of a typical type ab RR Lyrae variable, the
star's radial velocity varies by $\sim\pm50$ km/s about the systemic velocity
\citep{lay94}. To avoid excessive broadening of the spectral lines by
the changing velocity, the integration times of the spectroscopic
observations should be kept under 30 minutes, which is equivalent to
$\la 5\%$ of the pulsation cycle.  Given the faintness of the program
stars and these short exposure times, a very large telescope was
required for this project.  Our observations were obtained with the
Focal Reducer/Low Dispersion Spectrograph (FORS2) at the Very Large
Telescope - Yepun (VLT-UT4), European Southern Observatory (ESO),
Paranal, Chile, in service mode, during 18 nights between June and
August 2002.  We used grism 600B and a slit width of $1\arcsec$, which
yielded a resolution of $\sim6$ {\AA } and a dispersion of 1.20
\AA/pix.  The spectra were centered at 4650 \AA  and span the spectral
range from 3400 to 6300 \AA.

The service mode FORS calibration plan provides daily sets of bias and
flat field frames for the data reduction. For technical reasons
pertaining to this spectrograph, it is not possible to take wavelength
calibration exposures before and after an observation. Calibration
exposures can be taken only during daytime and with the telescope
pointing at the zenith.  Each day our targets were observed, one
daytime exposure of a He + HgCd lamp was made for wavelength
calibration. Instrument flexures are expected to be small , $<1$
pixel, and to depend on zenith distance \citep[e.g.][]{gal01}. 
As we explain in detail
below, we corrected the zero point of the wavelength calibration by
measuring several sky emission lines that were present in our spectra.

The exposure times for our targets varied between 20 and 30 minutes,
split into two exposures of equal length.  Because of the change of
radial velocity with phase, we requested two spectra of each star be
taken on different nights. Consequently, we have two spectra at
different phases for most of our targets, which facilitates the
fitting of a radial velocity curve to the measurements.  Only one
observation was obtained for 2 of the stars.  A few observations were
made outside the minimum requirements of seeing and/or airmass that
were specified for our program, and an additional observation was then
scheduled.  After reduction, we discovered that the spectra taken
under relatively poor conditions were nonetheless useful, which
resulted in 3 spectra being available for 3 stars. Table~\ref{tab-log}
gives the log of the observations. The identification number of
the RR Lyrae stars is the same as in \citet{viv04}.  

In addition to our 16 program stars, we requested spectra of 4 radial
velocity standard stars, which are also standard stars for Layden's
(1994) pseudo-equivalent width system (see
\S4). Table~\ref{tab-standards} lists the observation log of the
standard stars and the adopted values of heliocentric radial velocity,
which were taken from the SIMBAD astronomical database\footnote{SIMBAD
is operated at CDS, Strasbourg, France}.  Because we were uncertain
about the optimum exposure time for these bright stars, we requested
two observations of different length. In most cases, we could use both
spectra because they had high S/N and were not saturated. A companion
sky spectrum for each standard star was taken by moving the slit off
the star and on to blank sky. These exposures of 300s yielded good S/N
spectra of the sky lines, which were used to check and to correct, if
necessary, the wavelength calibration. The sky lines were too weak in
the short exposures of the standard stars themselves to be useful.

\subsection{Data Reduction}

The image reduction and extraction of the spectra was made using standard
IRAF\footnote{IRAF is distributed by
the National Optical Astronomy Observatories,
which are operated by the Association of Universities for Research
in Astronomy, Inc., under cooperative agreement with the National
Science Foundation} routines. Special care was given to the wavelength
calibration. Each spectrum was individually calibrated using an order 3,
cubic spline function for the fit and 12-14 spectral lines. The rms of the
dispersion solution was usually 0.02-0.03 \AA.

To check the wavelength calibration, we measured several emission
lines of the night sky that were also present in the relatively long
exposures of the targets.  The reference wavelengths of these lines
were taken from the Keck telescope web
page\footnote{\url{http://www2.keck.hawaii.edu/inst/lris/skyline.html}}.
Six sky lines could be measured in these spectra, including the strong
[OI] at 5577.338 {\AA }, and Na I at 5889.950 \AA.  The central
wavelengths of these lines were measured by fitting a Gaussian
profile. The mean deviations of sky lines from their reference
wavelengths varied from star to star and from -1.09 {\AA } to +0.05
\AA. Following \citet{gal01}, these deviations were used to 
correct the zero point of the
wavelength scale, which were applied to each spectrum by modifying the
header parameter CRVAL1, the starting wavelength.  Corrections were
applied to both our program stars and the radial velocity standards.
The standard deviation of the mean offset of the 6 sky lines from
their reference wavelengths was typically of 0.16 \AA, which is
equivalent to about 10 km/s. This value is included later in the
calculation of the error in the radial velocities.
 
Figure~\ref{fig-shift} shows that there is indeed a relation between
the amount of the shift of the sky lines and the zenith angle of the
telescope, which suggests that instrument flexure is to blame for at
least some of the zero-point offsets.  Other authors
\citep[eg.][]{tol00} have chosen to obtain the dispersion solution,
and not only the zero point correction, of FORS spectra by using sky
lines exclusively. This technique, which works well for red spectra
where there are numerous sky lines at all wavelengths, is
inappropriate for the few sky lines in our spectra.  The daytime lamp
spectra are used here to define the dispersion relation, but the zero
point of the calibration is corrected by measuring the wavelengths of
the sky lines.

There is still the possibility that the stellar spectra will be
displaced relative to the wavelength calibration because
the star is poorly placed in the relatively wide slit, which of course
has no effect on the sky lines.  This is potentially a problem for
all low resolution spectra where the projected slit width corresponds
to a large number of km/s.  During our observations, the star images
were seldom smaller than the slit width because the seeing varied from
$0\farcs 8$ to $1\farcs 7$ (Table~\ref{tab-log}), 
with an average value of $1\farcs 1$.
This probably smeared out the effects of guiding errors on the stellar
spectra, but this can be only checked by comparing the stellar velocities
between themselves and with published values (see below).

Because we split in two the exposure time of each program star, we
extracted, wavelength calibrated and corrected each observation
individually. Then, the two spectra were summed (using the IRAF task
{\sl scombine}) to improve the S/N.  Figure~\ref{fig-spectra} shows a
few examples of the reduced spectra of the RR Lyrae stars. Most of the
spectra have excellent S/N ($> 25$ at the Ca II H and K lines).  The
few spectra with worse S/N (bottom panel of Figure~\ref{fig-spectra})
are still good enough to measure the star's radial velocity.

\section{RADIAL VELOCITIES}

The radial velocities of the stars were measured with the IRAF task
{\sl fxcor}, which performs Fourier cross-correlation between a
program star and a radial velocity standard star.  The wavelength
range for the cross-correlation was from 3800 to 5200 \AA. As expected
with the high S/N of our spectra, the correlation peaks were well
defined in all cases.  They were, however, broad because the
correlations are dominated by the Balmer lines, which are wide
features in these stars with A-F type spectra.

It is highly desirable that the cross-correlation is made with 
a template that has a spectral type similar to the target's. For this
reason, we decided not to use the spectra of the standard star HD
97783 because none of our targets have such a late spectral type.
Each spectrum of a RR Lyrae star was therefore correlated with the 5
spectra of the 3 remaining standards.

While {\sl fxcor} returns a value for the standard deviation of the
cross-correlation ($\sigma_{cc}$), this is not a true measure of the
uncertainty because it does not take into account the zero-point
uncertainty in the wavelength calibrations of the program star and the
standard star.  The sigmas of these zero-points were therefore added
in quadrature to $\sigma_{cc}$.  Then using these adjusted values of
$\sigma_{cc}$ as weights in the standard way, we calculated the
weighted mean of the 5 velocities returned by the cross-correlations
and the standard deviation of this mean.  After correction for the
Earth's motion, these values were adopted as the heliocentric radial
velocity ($V_r$) and its standard deviation ($\sigma_r$), which are
listed in Table~\ref{tab-log}.

In order to determine the systemic velocity ($V_\gamma$), each star
was fitted with the radial velocity curve of the well-studied RR Lyrae
star X Aretis.  We used Layden's (1994) parameterization of the
velocity curve that \citet{oke66} measured from the H$\gamma$ line.  As
demonstrated by \citet{oke66} and other observers, the velocity curves of
type ab RR Lyrae variables that are measured from the Balmer lines
have larger amplitudes and are more discontinuous during the rise to
maximum light than the ones measured from the weaker metal lines.  The
Balmer line curves are appropriate for the observations made here, and
the good agreement that \citet{lay94} found between his results and
values of systemic velocity that are based on measurements of the
metal lines provides confidence in this method. In
Figure~\ref{fig-rv}, we show the fit obtained for 3 RR Lyrae stars in
our sample with Layden's (1994) curve. They are examples of a very
good fit (Star \#360), an average fit (Star \#244), and a poor fit
(Star \#409).  Because the form of the discontinuity near
maximum light is not the same in all type ab variables, this region is
best avoided when determining the systemic velocity.  Consequently,
observations at phases less than 0.1 or greater than 0.85 were
excluded from our fits of the velocity curves.  The top and middle
plots in Figure~\ref{fig-rv} contain observations that were excluded
from the fits for this reason (in each case the point to the far
right).  It is important to note that while these points and a few
additional observations of other stars are not useful for fitting the
velocity curves, they are, as expected, within the extremes of the
velocity curve.  This would not be the case if our observations were
plagued by very large errors stemming, for example, from very poor
guiding (see above).

Due to the variation in the values of $\sigma_r$ (see Table~\ref{tab-log}), we
weighted the values of $V_r$ when fitting the velocity curves.  For X
Ari, the systemic velocity ($V_\gamma$) occurs at phase 0.50, and
this phase was adopted for all the program stars.  Our estimation of
the error in the systemic velocity ($\sigma_\gamma$) includes the
uncertainty in this value.  Table~\ref{tab-rv} lists the number of
observations that were used in the fits, the values of $V_\gamma$,
and the rms of the fitted points (when $N\geq 2$).  The large range of
the rms values is not surprising given the small N.  Because the
average of these values (16 km/s) is very similar to $\sigma_r$, the
velocity curve of X Ari appears to be a reasonable model for the
program stars.

The errors in the values of $V_\gamma$ were estimated in the
following way.  Most of the observations used in the fits were made in
the phase interval 0.1 to 0.77, where Layden's (1994) curve is a
straight line.  The remaining observations were made between phases
0.77 and 0.85 where the velocity remains nearly constant.  For
estimating the errors, these points were considered to lie on the same
line as the others at a phase of 0.77.  The error in $V_\gamma$ that
is determined from one value of $V_r$ depends of course on
$\sigma_r$, but also on the uncertainty in the adopted velocity curve
as well.  This has been approximated by considering the likely uncertainties
in the phase where the velocity curve passes through the systemic
velocity and in the slope of the velocity curve, given the fact that
there are star-to-star variations in the amplitudes of the velocity
curves.  The type ab star SU Draconis was observed by \citet{oke62}
in essentially the same manner that \citet{oke66}
observed X Ari.  The H$\gamma$ velocity curve of SU Dra is linear
between phases 0.0 to 0.8, but passes through systemic velocity at
phase 0.45 instead of 0.50 as in the case of X Ari.  To be
conservative, we adopted 0.1 as the 1$\sigma$ uncertainty in the
phase of the occurrence of systemic velocity, which also includes some
allowance for the errors in the determination of the phases of our
observations.  The high-quality velocity curves of 22 type ab
variables that were discussed by \citet{liu91}, which were
not determined from measurements of the Balmer lines but from metal
lines, indicate that the velocity amplitude of X Ari lies near the
middle of the range.  It is likely that this is also true of the
amplitude of the velocity curve given by its Balmer lines.  In this
connection, it is important to note that the amplitudes of the
velocity curves and the light curves are tightly correlated 
\citep[e.g.][]{liu91} and that the light variation of 
X Ari is only sightly larger
than the average amplitude of our program stars.  Over the phase
interval 0.1 to 0.77, the Layden's (1994) curve for X Ari has a slope
of 119.5 km/s per unit phase, and from the above considerations,
we estimate 23.9 km/s as the $1\sigma$ variation in this number stemming
from the star-to-star variations in velocity amplitude.  In the
following equation for the error in $V_\gamma$, $\Delta\phi$ is the
difference in phase between 0.50 and the phase of the observation.
The second and third terms take into account the uncertainties in the
phase of the systemic velocity and the slope of the velocity curve,
respectively.

\begin{equation}
\sigma_\gamma^2=\sigma_r^2 + (119.5*0.1)^2 + (23.9*\Delta\phi)^2
\label{eq-sigma}
\end{equation}

This equation was used to calculate $\sigma_\gamma$ for each
observation used to measure $V_\gamma$.  If two observations were
used to determine $V_\gamma$, the final value was
calculated by combining the values of $\sigma_\gamma$ from the
individual measurements in the usual manner for calculating the $\sigma$
of a weighted mean value.  Table~\ref{tab-rv} lists the values of
$\sigma_\gamma$ that we obtained.

We have a few external checks on the accuracy of our radial
velocities.  When cross-correlated with each other, the 7 spectra of
the 4 radial velocity standards yield, to within the errors, the
standard values of their velocities.  Because the spectra of these
stars were taken over a range of zenith angles ($39-47\degr$), this
comparison is also a check on our method to calibrate the zero points
of the wavelength calibrations.  For a separate project we obtained a
spectrum of an RR Lyrae star in the globular cluster Palomar 5 (Pal 5
V2; QUEST ID = 400) during the same observing period and with the same
instrumental setup as the observations reported here.  Exactly the
same measuring techniques yield $V_r = -63 \pm 15$ km/s. Since V2 is a
type c RR Lyrae variable, it is expected to have a much smaller
velocity amplitude than type ab variables and at most small
differences are expected between the Balmer-line and the metal-line
velocity curves \citep{tif58}.  The velocity
curve of X Ari is clearly inappropriate for it, and we considered
instead the velocity curves that \citet{liu89}
and \citet{jon88} measured from metal lines
for the type c variables TV Boo, T Sex, and DH Peg.  The shapes of
these curves are quite similar, and they indicate that phase of
systemic velocity is near 0.34.  Our observation of V2 in Pal 5 was
made at a phase of 0.37; hence to obtain $V_\gamma$, it is necessary
to make only a very small correction to $V_r$.  Using the velocity
curves of the above 3 variables as templates, we obtain $V_\gamma =
-65 \pm 16$ km/s for V2, and the error includes an estimate of the
uncertainty due to differences in the velocity curves.  The
heliocentric velocity of Pal 5 ($-58.7 \pm 0.2 $ km/s) and its
internal velocity dispersion ($1.1 \pm 0.2$) are very precisely known
from the work of \citet{ode02}.  To within the errors, our
measurement of V2 is consistent with membership in Pal 5, which on the
basis of position on the sky and distance from the Sun is a near
certainty.

The distribution of radial velocities of the observed sample of RR
Lyrae stars is very narrow (Figure~\ref{fig-rvhist}), which is a clear
sign that it is not a random sample of halo stars, but as expected a
coherent group in velocity space. For comparison, we estimated the
mean radial velocity of halo stars in this part of the sky
($l=340\degr$, $b=+56\degr$) by assuming a non-rotating halo and
following the procedure described in \citet{yan03}. The resulting
distribution of random halo stars is shown in Figure~\ref{fig-rvhist}
as a dashed Gaussian curve with $\sigma=100$ km/s.  The mean velocity
of the 16 RR Lyrae stars is 13 km/s with a standard deviation of 62
km/s (after subtracting in quadrature our observational
errors). However, two of the stars (\#333 and \#337) have velocities
far from the rest of the group.  According to \citet{vz04}, if the Sgr
stream lies within a smooth distribution of halo stars following a
$r^{-3}$ power-law, we expect 1-2 halo RR Lyrae stars in this volume
of space. Thus, the two outliers may belong to the general halo
population (but see \S5).  Removing these two stars, we obtain a
mean velocity for the group of $33\pm 8$ km/s with a standard
deviation of only 25 km/s. We will show below that these results agree
qualitatively with the models of the Sgr tidal stream, but also that
the velocity and the spatial distributions of the RR Lyrae variables
are very useful for distinguishing between different models, none of
which match these observations in full detail.

The radial velocity measurements by other authors of stars in this Sgr
stream agree with our observations. There are 4 red giant stars from
the Spaghetti Survey that lie close on the sky to our region
\citep{doh01}.  The velocity distribution of those stars is also shown
in Figure~\ref{fig-rvhist}. This group has a mean velocity of $42\pm
16$ km/s and a standard deviation of 32 km/s.  The recent work of
\citet{maj04} provides velocities for a large number of M giants in
the stream, most of them belonging to the trailing tidal tail.  There
are, however, 3 stars in their sample that lie close to the region on
the sky that was observed by the QUEST survey.  These stars have a
mean velocity of 64 km/s ($\sigma=43$ km/s), which is consistent with
the velocity distribution that we have found for the RR Lyrae
variables.  According to \citet{maj03,maj04}, the mean distance from
the Sun of this part of the Sgr stream, in particular these 3 stars,
is $\sim40$ kpc, which is considerably smaller than the distances of
the RR Lyrae variables ($\sim54$ kpc).  This castes some doubt on
their membership in the same stream and/or the compatibility of the
distance scales for the two types of stars.  The blue horizontal
branch stars that have been identified by the Sloan Digital Sky Survey
(SDSS) in the northern stream form a broad clump at $\sim 50$ kpc from
the Sun and at $V_{los} \sim -10$ km/s (see Fig. 14 in
\citealt{sir04}), which corresponds to $V_{r} \sim +10$ km/s.  The
SDSS survey has also identified candidate RR Lyrae variables in this
part of the sky, and measurements of $V_{r}$ for a subsample of these,
without correction for the variation in $V_{r}$ as a function of phase,
has a broad distribution with a peak at $V_{r}\sim40$ km/s \citep{ive03}.
Both of these measurements
from the SDSS survey are consistent with what we find from a smaller, but more
precisely measured sample of stars.

To investigate in more detail the velocity distribution of the leading
Sgr stream, we have computed for our sample of RR Lyrae variables
the longitude ($\Lambda_{\sun}$) and latitude
($B_{\sun}$) in the Sgr orbital plane (\citealt{maj03} and 
\url{www.astro.virginia.edu/{\tiny$\sim$}srm4n/Sgr}) and
the Galactocentric Standard of Rest radial velocity ($V_{GSR}$), which
required removing from $V_{\gamma}$ the contributions from the Sun's
peculiar motion and the rotation of the Local Standard of Rest (see
Table~\ref{tab-rv}). We
used the same values for these corrections as did \citet{maj04} so
that a direct comparison can be made with their results for the
trailing Sgr stream.  \citet{maj04} found that the dispersion in
$V_{GSR}$ varies as a function of $\Lambda_{\sun}$ from
$10.4\pm1.3$km/s ($25\degr<\Lambda_{\sun}<90\degr$) to $12.3\pm1.3$
($90\degr<\Lambda_{\sun}<150\degr$)\footnote{$\Lambda_{\sun}=0\degr$ 
is the Sgr core}.  
They obtained these values after fitting a quadratic equation
in $\Lambda_{\sun}$ to the run of $V_{GSR}$ with $\Lambda_{\sun}$.  Among
our sample of RR Lyrae variables, there is no evidence for more than a
linear dependence, which may be a consequence of our larger measuring
errors and the fact that our sample spans only $28\degr$ in
$\Lambda_{\sun}$.  In addition, there is a strong correlation between
$\Lambda_{\sun}$ and $B_{\sun}$ in the sample (see Table~\ref{tab-rv}),
which is due to the inclination of the Sgr plane to the declination
band of the QUEST survey.  A variation of $V_{GSR}$ with $B_{\sun}$ may
therefore skew the one with $\Lambda_{\sun}$.  With the very deviant
stars \#333 and \#337 removed from the sample, a straight line fit to
the ($\Lambda_{\sun}$,$V_{GSR}$) pairs yields a dispersion of 30 km/s.
After subtracting in quadrature the average value of $\sigma_{\gamma}$
(16.6 km/s), we obtain $25\pm8$ km/s for the intrinsic velocity
dispersion.  
To compute the uncertainty in this value, we followed the usual procedure
for the propagation of errors and the usual practice that the uncertainty
in the $\sigma$ given by the fit is $\sigma/\sqrt{2N}$, $N=$ number of
observations (14).  For the random error in the average $\sigma_{\gamma}$,
we adopted 5 km/s on the basis that this average is unlikely to exceed by
much our estimate of $\sigma{\gamma}$ for one observation.
The velocity dispersion of
this sample of RR Lyrae variables in the leading Sgr stream is
therefore significantly larger than the ones that \citet{maj04} found
for the M giants in the trailing stream.  Some part of this difference
may be related to the fact that the RR Lyrae variables are members of
an older stellar population than the one producing the M giants, which is
probably significantly younger than 5 Gyrs \citep{maj03}.  The RR
Lyrae sample may therefore contain some stars that were released from
Sgr long before the release of the M giant population.

\section{METAL ABUNDANCES}

The metal abundances ([Fe/H]) of the RR Lyrae variables have been
measured using the variation of the $\Delta$S technique that was
pioneered by \citet{fre75}. In this method, the pseudo-equivalent
width of the CaII K line, W(K), is plotted against the mean of the
pseudo-equivalent widths of the Balmer lines of hydrogen, W(H).
Except during a small period of time on the rising branch, the
variations in these parameters with phase define tight sequences that
are essentially identical for stars of the same [Fe/H].  The curves
for stars of different [Fe/H] are systematically offset from one
another, which after suitable calibration with stars of known
abundance, allows one to measure [Fe/H] from a low resolution
spectrogram. This technique yields abundances that are erroneously too
low when it is applied to spectrograms taken during the rapid increase
in effective gravity on the rising branch of the light curve when
shock waves produce emission in the cores of the Balmer lines.  The
magnitude of this effect is largest in the largest amplitude type ab
variables and can vary from cycle to cycle (see \citealt{smi95} for a
review).  Two of the spectra (\#244 on 2002-06-17 and \#360 on
2002-08-10) have oddly shaped Balmer line profiles, which is a sign
that they have been affected by this phenomenon \citep[e.g.][]{lay94}.
Consequently, we have discarded the values of W(K) and W(H) that were
derived from them.  The phases of a few other observations indicate
that they were made on the rising branch, but neither the spectra nor
the values of [Fe/H] that were derived from W(K) and W(H) are unusual.

\citet{lay93,lay94} has calibrated this technique on the widely used
\citet{zin84} metallicity scale for globular clusters.  Although this
scale is two decades old, over the range $-1.7<$[Fe/H]$<-1$ it is in good
agreement with the [Fe/H]$_{\rm II}$ scale that \citet{kra03}
recently derived from high-dispersion spectroscopy of red
giants.  At larger and smaller abundances, the differences between the
scales are small ($\le 0.3$ dex) and depend to some extent on the way
the high dispersion measurements are extrapolated.  We have followed
very closely Layden's analysis by using his computer program to measure the
pseudo-equivalent widths and by observing standard stars from his list
of equivalent width standards so that our measurements could be
transformed to his system.  We also used Layden's method for
correcting W(K) for interstellar absorption, which is small because of
the high galactic latitudes of our stars.  Since we did not measure RR
Lyrae variables of known abundance, we did not attempt to correct the
pseudo-equivalent width of H$\beta$ for metal-line absorption as did
\citet{lay94}.  This is a minor effect, and the results \citet{lay93}
obtained with and without this correction have nearly identical
precisions.

Our measurements of W(H), the mean value of the widths of the
H$\beta$, H$\gamma$, and H$\delta$ lines, and W(K) are listed in 
Table~\ref{tab-log}
and plotted in Figure~\ref{fig-HK}.  W(H) and W(K) were not measured for the
observations of stars \#294 on 2002-08-08 and \#379 on 2002-07-06
because these spectra are too noisy to yield reliable results.
Because W(H) and W(K) vary with phase, we cannot use most of the
multiple observations of the same stars to estimate the observational
errors.  Three stars were observed at two phases that differ by $<0.1$, and
the average deviations of the measurements of W(H) and W(K) from their
mean values are $\pm0.22$ and $\pm0.17$, respectively.  Because some
fraction of this variation may be due to the difference in phase,
these values are probably upper limits on the true errors.

During their light cycles, type ab variables describe loops in the
W(H)-W(K) plane that collapse to approximately straight lines once the
period of high effective gravity on the rising branch is excluded.
Layden's (1993, 1994) metallicity calibration consists of a family of
straight lines that he fit to observations of stars of known [Fe/H].
We have compared our observations with a few of these lines in 
Figure~\ref{fig-HK},
where one can see that the lines connecting two observations of the
the same star that differ by a large amounts in W(H) generally follow
Layden's lines of constant [Fe/H].  Several of these measurements are
systematically offset from one another in the sense expected of stars
that differ in [Fe/H].  This and the large range in W(K) that is
spanned at roughly constant W(H) is evidence for a real metallicity
variation among this sample of stars.

The [Fe/H] values (see Table~\ref{tab-log}) were obtained from the lines in
Layden's (1993) family that passed through each W(H),W(K) measurement.
More than one [Fe/H] measurement was obtained for 13 of the stars, and
the deviations of these measurements from their mean values average
0.08 dex.  The errors in Layden's calibration of the
\citeauthor{fre75} technique appear to be the major source affecting
the external precision of our measurements.  Layden's (1994)
measurements with spectrograms similar to ours in resolution and
signal to noise had precisions of 0.15-0.20 dex when compared to
other [Fe/H] measurements in the literature.  The external precisions
of our measurements are probably similar to his, although we do not
have an independent determination. Average metallicities for each star
are given in Table~\ref{tab-rv}.

Figure~\ref{fig-fehhist} shows the histogram of the mean [Fe/H] values
for the 16 RR Lyrae variables.  The distribution is quite narrow
($\sigma=0.21$ dex) around a mean value of $-1.77$. The
removal of the two stars that may not belong to the Sgr stream change
these numbers slightly ($\langle{\rm [Fe/H]}\rangle = -1.76$, $\sigma
= 0.22$).  The distribution of [Fe/H] that we find agrees with the
metallicities of the 4 red giant stars measured by the Spaghetti
Survey (shaded region in Figure~\ref{fig-fehhist}).  
The dispersion of the RR Lyrae variables is
larger that expected from the measuring errors alone, which is
evidence for a real abundance range.  It is substantially smaller than
the dispersion in the Milky Way's halo ($\langle{\rm
[Fe/H]}\rangle_{\rm halo} =-1.6$; $\sigma_{\rm halo}=0.4$ dex;
\citealt{kin00}) and is smaller than the ones found in the dSph
galaxies in the Local Group \citep[0.3-0.6 dex;][]{mat98}.  However,
the measurements for these galaxies pertain to their entire stellar
populations, which in most cases span a much larger range in age, and
therefore possibly [Fe/H], than their RR Lyrae variables.

The main body of Sgr has, in fact, a well-defined age-metallicity
relation.  The $\langle{\rm [Fe/H]}\rangle$ found here for the RR
Lyrae variables in the leading Sgr stream is consistent with this
relationship \citep{lay00} if, as expected, they
are coeval with the oldest stellar population in the main body.  The
red giant sample from the Spaghetti Survey is, however, more metal
poor by $\sim 0.5$ dex than the average abundance of the red giants in
the main body \citep{doh01}.  Since the streams are composed of the
stars that were once at the periphery of the parent galaxy,
\citeauthor{doh01} speculated that this offset in [Fe/H] could be a sign
that a strong radial gradient in [Fe/H] existed in the dSph galaxy
prior to the formation of the streams.  There is no need of gradient
to explain the properties of the RR Lyrae variables, and it is
difficult to reconcile a large one with the recent discovery of many
metal-rich M giants in the streams \citep{maj03}.

The absolute magnitudes of the RR Lyrae variables have a mild
dependence on metallicity, and in their discussion of the QUEST
survey, \citet{vz04} assumed [Fe/H]$=-1.6$ when computing the absolute
magnitudes of the stars and their positions in the Milky Way.  The use
of our measured values of [Fe/H] in place of this average value
should, in principle, produce more accurate results.  Table 3 lists
the heliocentric distances (r$_\sun$) obtained using these
measurements, the interstellar extinctions in \citet{vz04}, and the
relation given in \citet{dem00}:

\begin{equation}
M_V (RR) = 0.22\mbox{[Fe/H]} + 0.90
\label{eq-mvfe}
\end{equation}
 
Because the mean [Fe/H] of the stars is only 0.17 dex lower than the
previously assumed value, this revision increases the distance to the
stream by only $\sim 1 $ kpc.  Excluding stars \#333 and \#337 on the
basis of their radial velocities, the sample has a mean of $r_\sun$ of
$53.8 \pm 1.3$ kpc and a standard deviation of 4.8 kpc.  Since we
purposely picked stars that span the range in magnitude of the stream,
the true distribution is probably more sharply peaked than this
sample.

\section{COMPARISON WITH RECENT MODELS}

Several numerical models of the disruption of Sgr are available in the
literature, and in general, there is good agreement between the
location of the multiple observations of the Sgr streams and the
predictions of these models, which are being refined as more data
become available.  Detailed comparisons are made here between the
properties of the RR Lyrae stars in the northern stream and the most
recent theoretical models: \citet{hel04} and \citet{mar04}.  We
compare not only the location of the stream on the sky but also the
distances and radial velocities of the stars.  To do this comparison,
we extracted from the models all the particles (stars) located in the same
part of the sky as our RR Lyrae variables: $13\fh 0 < \alpha < 16\fh 0$
and $-2\fdg 2 < \delta < +0\fdg 1$.

\begin{figure}
\plotone{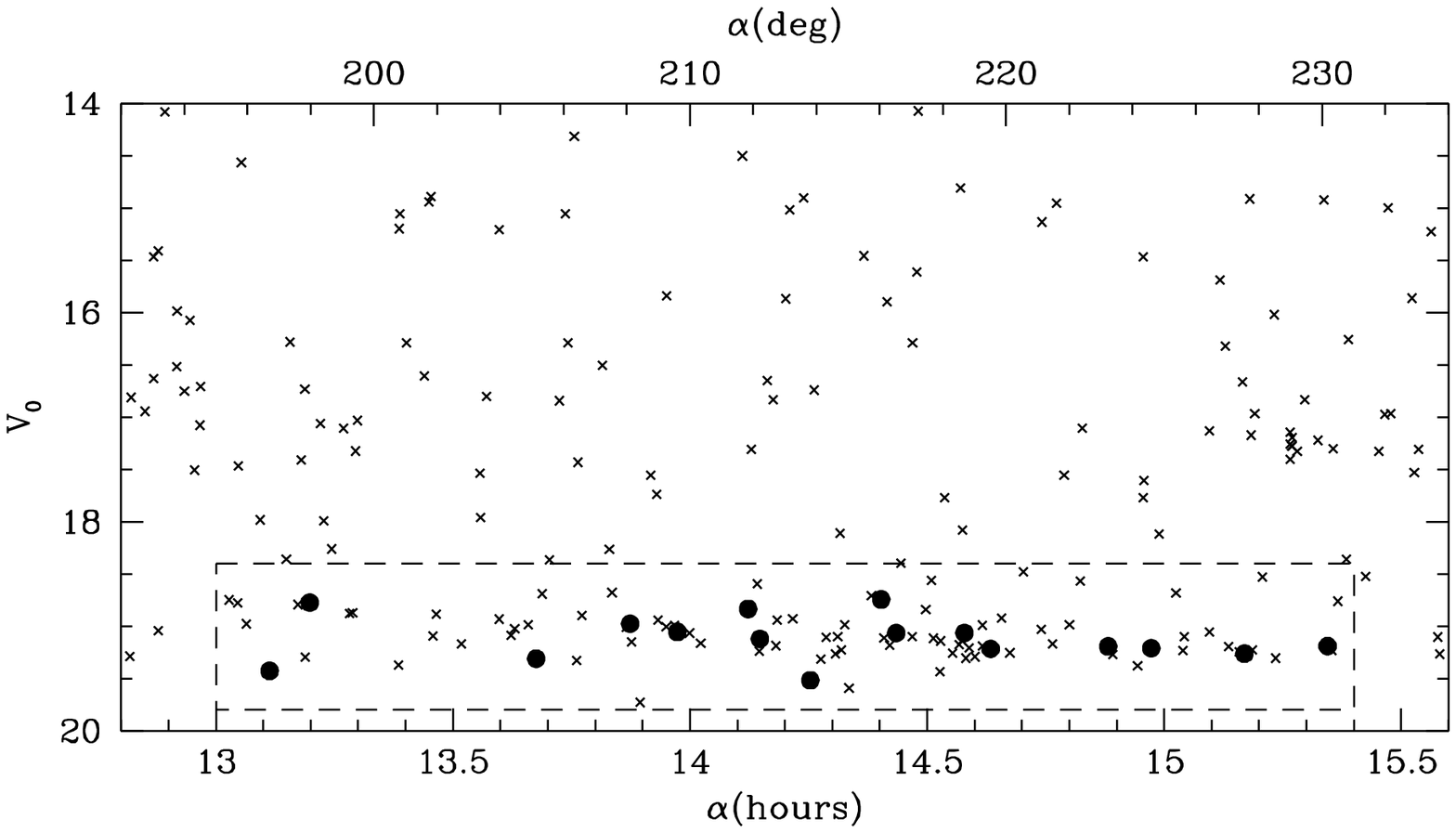}
\caption{Extinction corrected V magnitudes as a function of right
ascension for the subset of the QUEST RR Lyrae stars (crosses) with
$12\fh 8 < \alpha < 15\fh 6$. The region observed by the QUEST survey
consists of a long, but narrow ($2\fdg3$) strip of the sky centered on
$\delta=-1\fdg0$. The dashed lines enclose the stars that are
suspected to be part of the Sgr stream. Solid points indicate the
targets selected for spectroscopy at VLT. \label{fig-targets}}
\end{figure}

\begin{figure}
\plotone{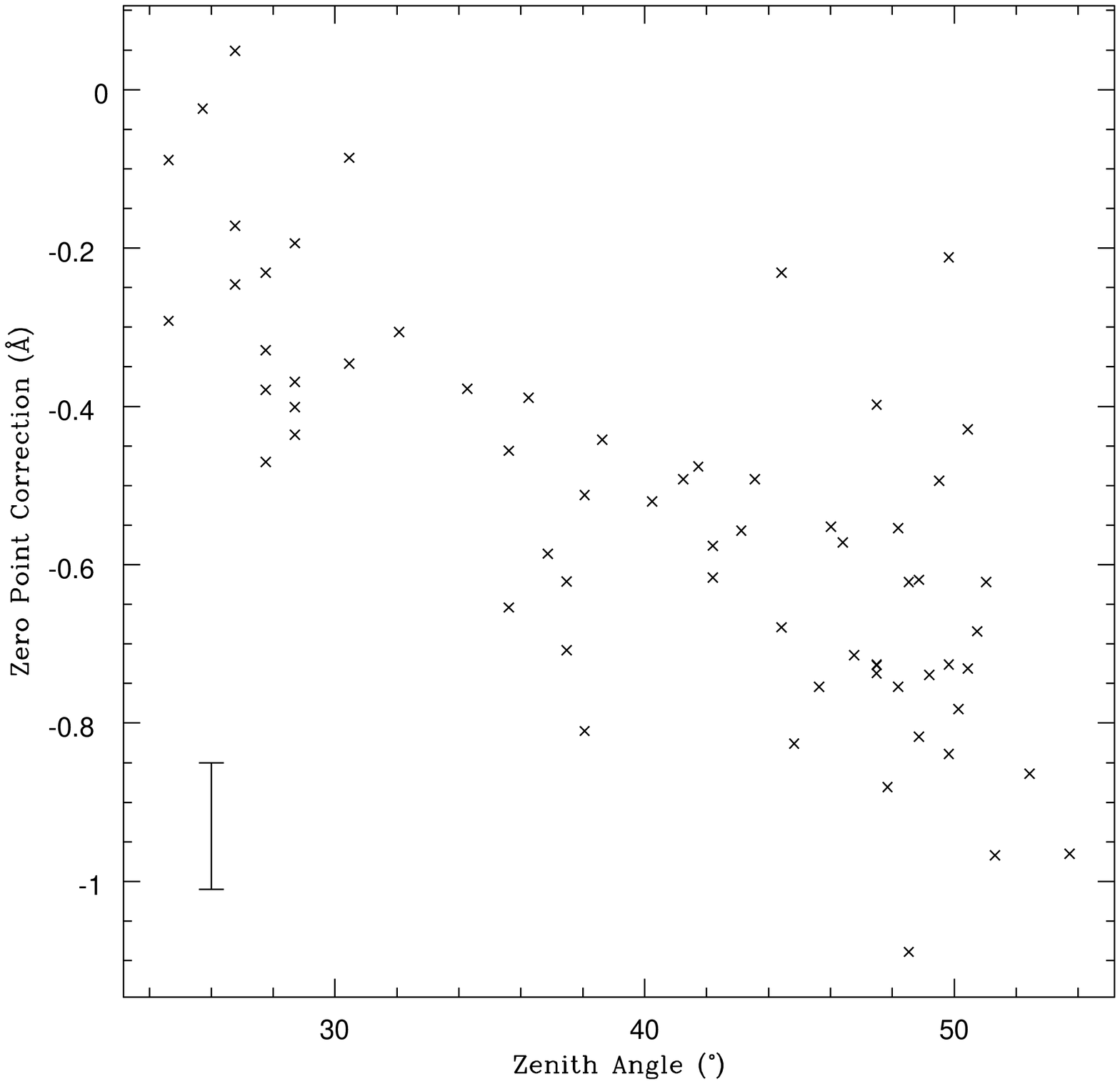}
\caption{The zero point corrections from the measurements of 6
emission sky lines in our spectra are plotted as a function of the
zenith angle of the observation. Note the correlation between these
quantities. The error bar in the lower left corner
indicates the typical standard deviation of the zero points.
\label{fig-shift}}
\end{figure}

\begin{figure}
\plotone{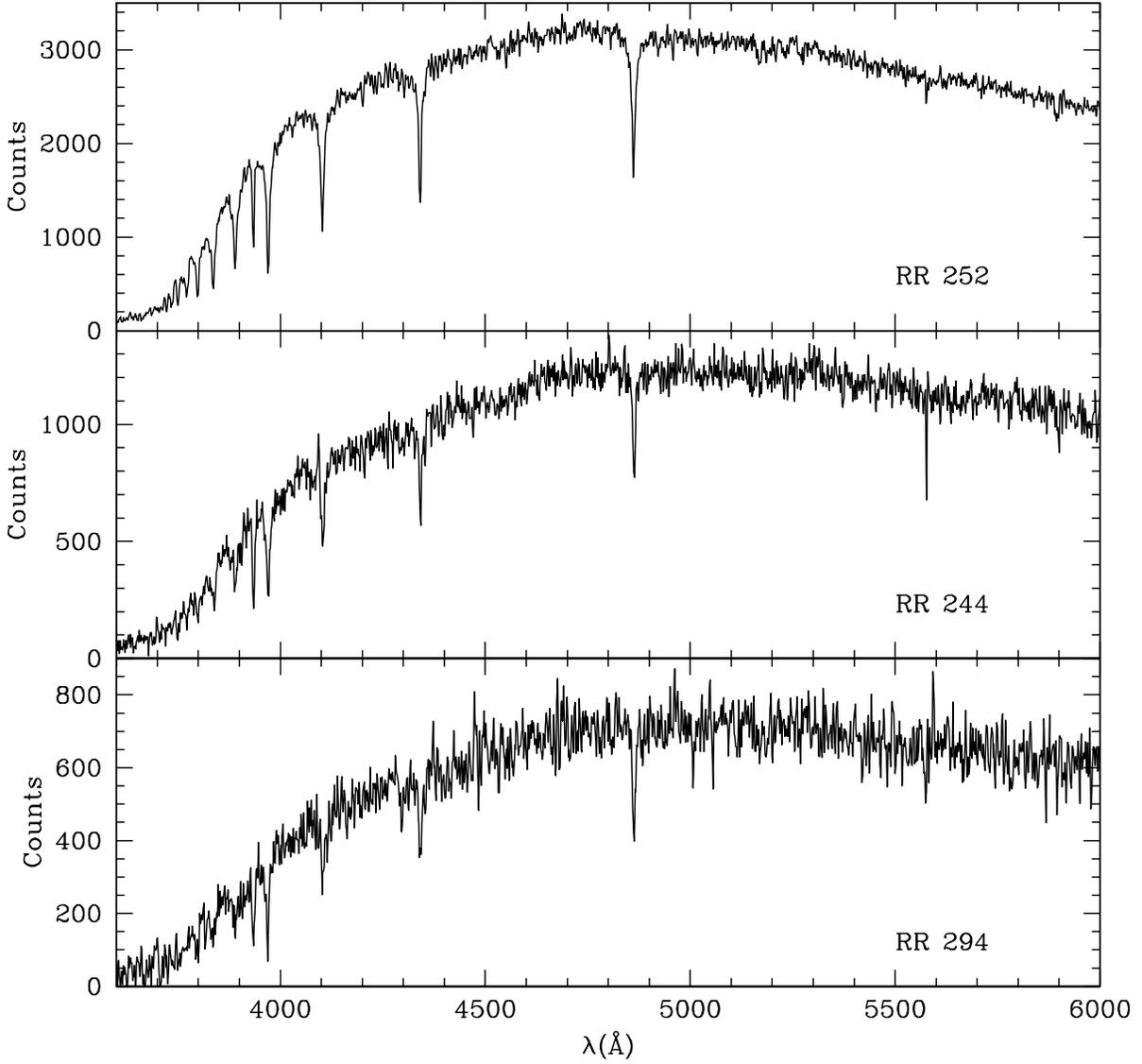}
\caption{Example of three of the VLT spectra of RR Lyrae stars. The bottom 
panel shows one of the spectra with the lowest S/N, which is still adequate
for measuring radial velocity.\label{fig-spectra}}
\end{figure}

\begin{figure}
\plotone{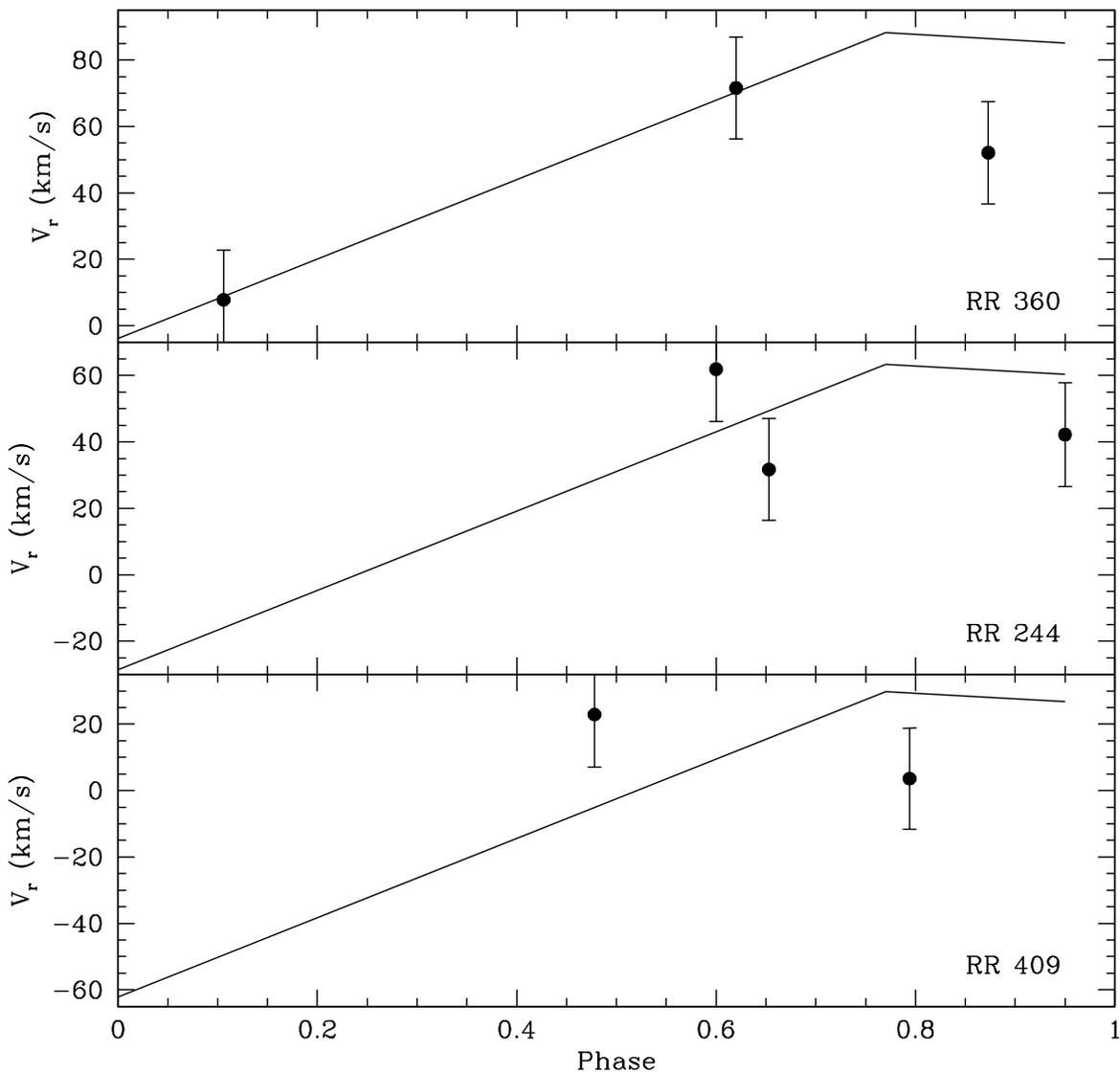}
\caption{Fits of the radial velocity template of X Ari (solid line) to
three stars in our sample. The solid points are the observational data.
The rightmost points in the top and middle panels were not used in the
fit because they lie too close to the discontinuity in the radial velocity
curve. We show an example of a very good fit (top), an average fit (middle)
and a poor one (bottom).
\label{fig-rv}}
\end{figure}

\begin{figure}
\plotone{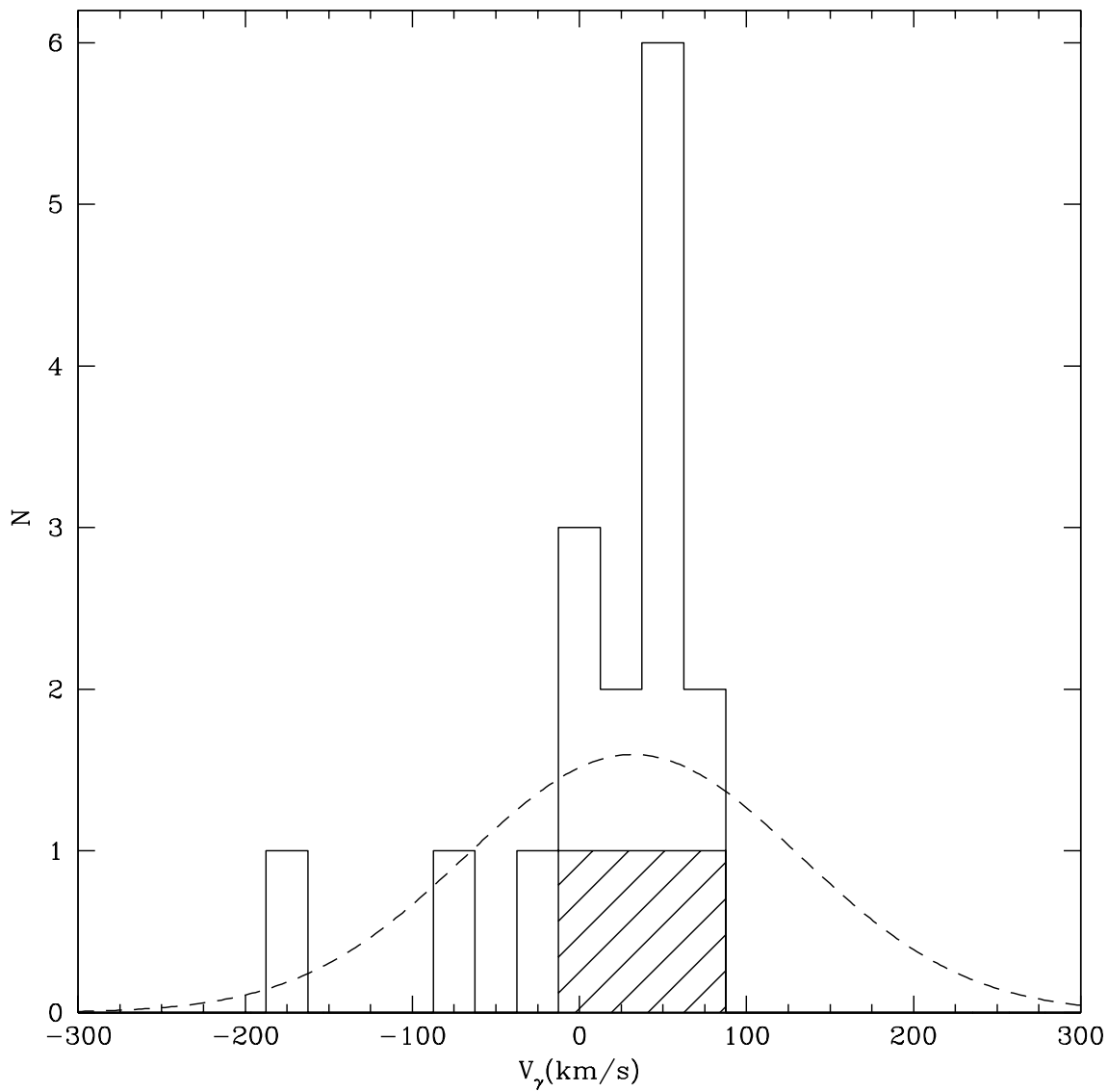}
\caption{Histogram of the heliocentric radial velocities of the 16 stars 
observed with VLT. For comparison we also show the distribution of the
red giant stars in the Spaghetti survey (shaded region). The dashed line 
shows the expected distribution of a random sample of halo stars in this
part of the sky (mean$=32$ km/s; $\sigma=100$ km/s).
\label{fig-rvhist}}
\end{figure}

\begin{figure}
\plotone{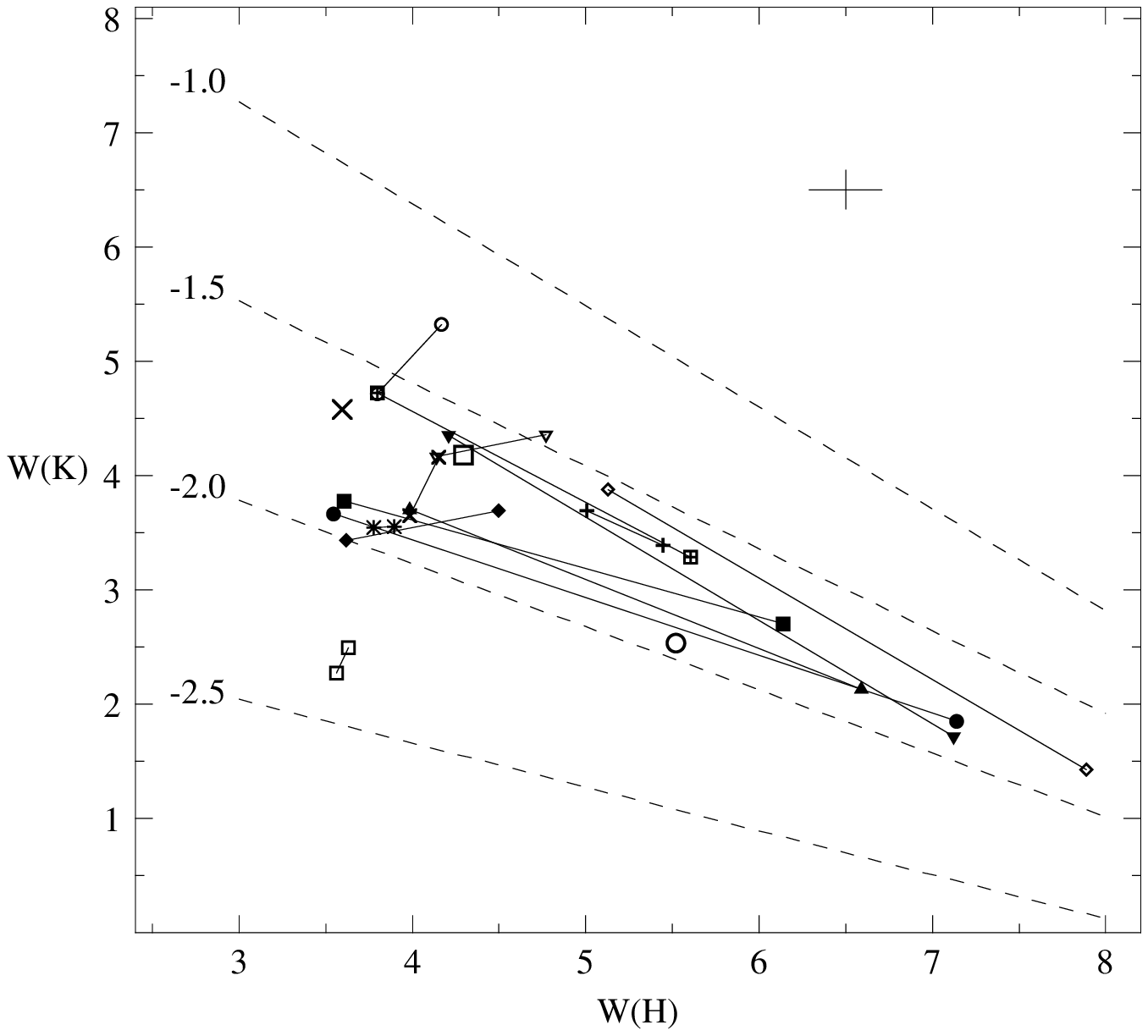}
\caption{W(K), the pseudo-equivalent width of the CaII K line
corrected for interstellar absorption, is plotted against W(H), the
mean of the pseudo-equivalent widths of the H$\beta$, H$\gamma$, and
H$\delta$ lines (both axes in \r{A}).  Solid lines connect the two
observations of the same star, which may differ by large amounts
because W(K) and W(H) vary with phase.  Each star is plotted with a
different symbol, and the largest ones denote the 3 stars that were
observed only once.  The dashed lines are the loci of stars that have
the indicated [Fe/H] values, according to the Layden (1993)
calibration.
\label{fig-HK}}
\end{figure}

\begin{figure}
\plotone{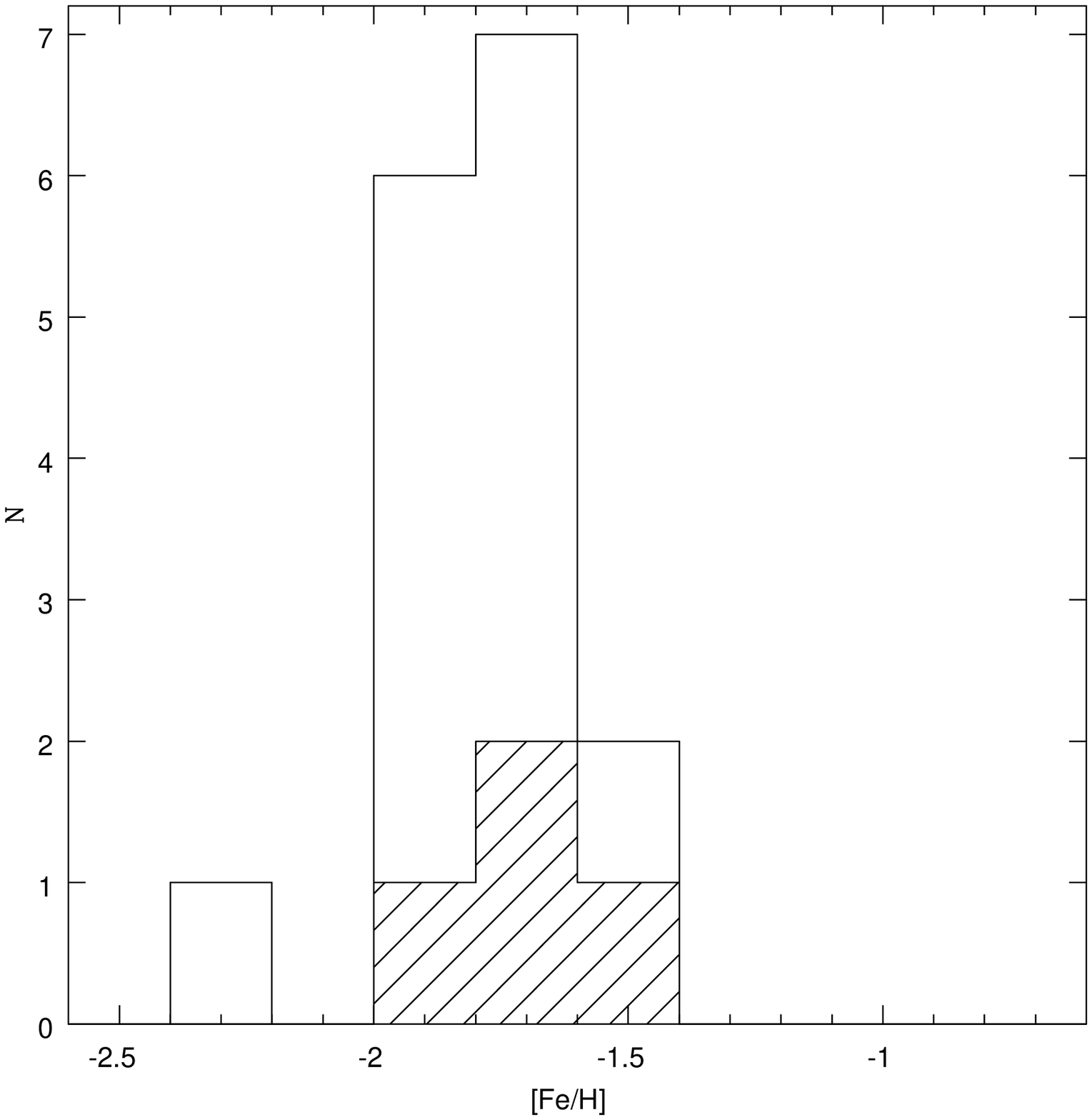}
\caption{Histogram of the metal abundances of the 16 RR Lyrae stars 
observed with VLT. For comparison we also show the distribution of the
red giant stars in the Spaghetti survey (shaded region).
\label{fig-fehhist}}
\end{figure}

\paragraph{The Size of the Stream.}
As discussed in \citet{viv01} and \citet{vz04} and shown in
Figure~\ref{fig-targets}, there is an over-density of RR Lyrae stars
spanning a very wide range in right ascension, $\sim36\degr$, from
$\alpha\simeq 195\degr$ to $232\degr$.  Our spectroscopic observations
confirm that this is the true size of the stream, for from one end to
the other, the RR Lyrae variables have a coherent velocity
distribution, with very little gradient in mean velocity.  Because RR
Lyrae variables are excellent standard candles, both the distance and
the width along the line of sight are reliably known in
this part of the sky.

\begin{figure}
\plotone{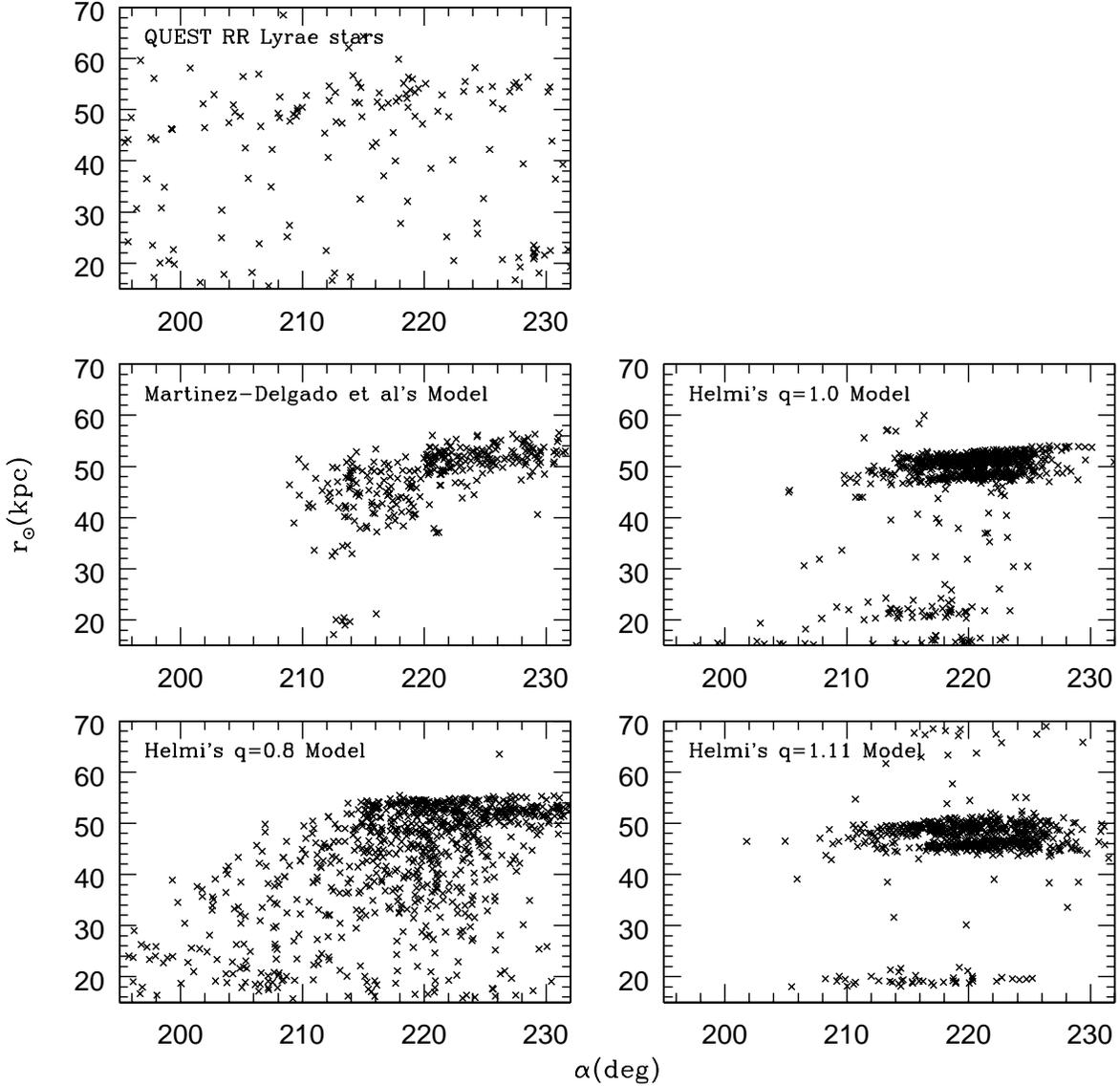}
\caption{The top panel plots the distances of the QUEST RR Lyrae stars
as a function of right ascension in the region of the Sgr stream.  The
stream is the concentration of stars at $\sim50$ kpc from the Sun. We
compare this distribution with the predictions of different
theoretical models of the disruption of Sgr \citep{mar04,hel04}.  The
plots on the lower left show the distribution of stream particles in
two models that assume oblate flattened dark matter halos for the
Milky Way. The plots on the right are for models that assume a
spherical halo and a prolate flattened one.
\label{fig-dist}}
\end{figure}

The top panel of Figure~\ref{fig-dist} is a plot of $r_\sun$ against
$\alpha$ for all the RR Lyrae stars in the QUEST survey that lie
within the range of $\alpha$ shown by the dashed rectangle in
Figure~\ref{fig-targets} and have $r_\sun \ge 15 $ kpc.  The number of
stars at $\sim50$ kpc is much larger than the expected number of
random halo stars \citep{viv01,vz04}. There are two other
over-densities of RR Lyrae stars which are described with detail in
\citet{vz04}. At $\alpha\sim230\degr$, $r_\sun \sim20$ kpc there are
variables in the globular cluster Pal 5 and in addition, ones that are
probably part of its tidal debris.  The small group of stars at
$\alpha<200\degr$ and $r_\sun \sim20$ kpc in the figure is the eastern
edge of a larger feature that has been called the {\sl ``$12\fh 4$
clump''} \citep{viv03,vz04}. Besides these three over-densities, the number and
distribution of the rest of the stars is compatible with them being
random halo stars.  The other panels in the Figure show the distance
distributions of the stars in the Sgr stream that are predicted by the
models of \citet{mar04} and \citet{hel04}.  The model of
Mart{\'\i}nez-Delgado et al. that is plotted assumed a flattening of
the dark matter halo of q=0.5.  They concluded that this model yielded
the best fit to the available observational data. Also plotted are
three different models by Helmi, each assuming a different shape for
the Milky Way's dark matter halo: an spherical halo (q=1.0), an oblate
flattened halo (q=0.8) and a prolate flattened one (q=1.11)

Each of these models fails to reproduce in detail the location of the
stream on the sky.  The highest concentration of the RR Lyrae stars
occurs at $\alpha\sim216\degr$, coincident with the Sgr plane defined 
by the M-giants \citep{maj03},
while all the models have their peaks at alpha~220.  
In addition, none of the models reproduces the length of the stream in
right ascension. There are
very few particles in each model that lie at distances of $\sim50$ kpc
with $\alpha<210\degr$, while there are many RR Lyrae stars with
$195\degr<\alpha <210\degr$ and $r_\sun\sim50$ kpc.

The region between $210\degr<\alpha<220\degr$ provides particularly
useful comparisons because there are clear differences between the
oblate models (\citeauthor{mar04} and \citeauthor{hel04}'s q=0.8
model) and the spherical and prolate ones, and because there is no
possible confusion in the QUEST data between the Sgr stream and the
other over-densities mentioned above. In this region, both oblate
models predict large widths in $r_\sun$, which are not seen in the
QUEST data. Within this span of $\alpha$, there are 34 RR Lyrae stars at
$r_\sun>45$ kpc and only 7 between $35<r_\sun<45$ kpc. As mentioned
above, this small number of relatively nearby stars is consistent with
their membership in the general halo population.  In the
Martinez-Delgado et al. model, there is a broad distribution of stream
particles between $\sim$35 and 55 kpc, with no strong
clustering. Helmi's oblate model is qualitatively similar in having a
broad distribution in distance, although it does show some
concentration at 50 kpc: $54\%$ of all the particles in the model have
$r_\sun>45$ kpc, while $35\%$ have $35<r_\sun<45$ kpc.  Since the
QUEST survey easily found the over-density at $\sim50$ kpc, it is
impossible that this survey missed an almost equally dense region at
smaller distances.

Both Helmi's spherical and prolate halo models predict a much narrower
stream, with dispersions in distance of only $\sim3$ kpc.  This is
much more consistent with the distribution of RR Lyrae variables than
the oblate models.  The spherical and prolate models are, however,
offset in $r_\sun$ by about 5 kpc ($\Delta(m-M)_o\sim0.2$) from the
concentration of the variables.  This is too large to be due to the
likely errors in the RR Lyrae distance scale alone \citep[see][]{cac03}.

\begin{figure}
\plotone{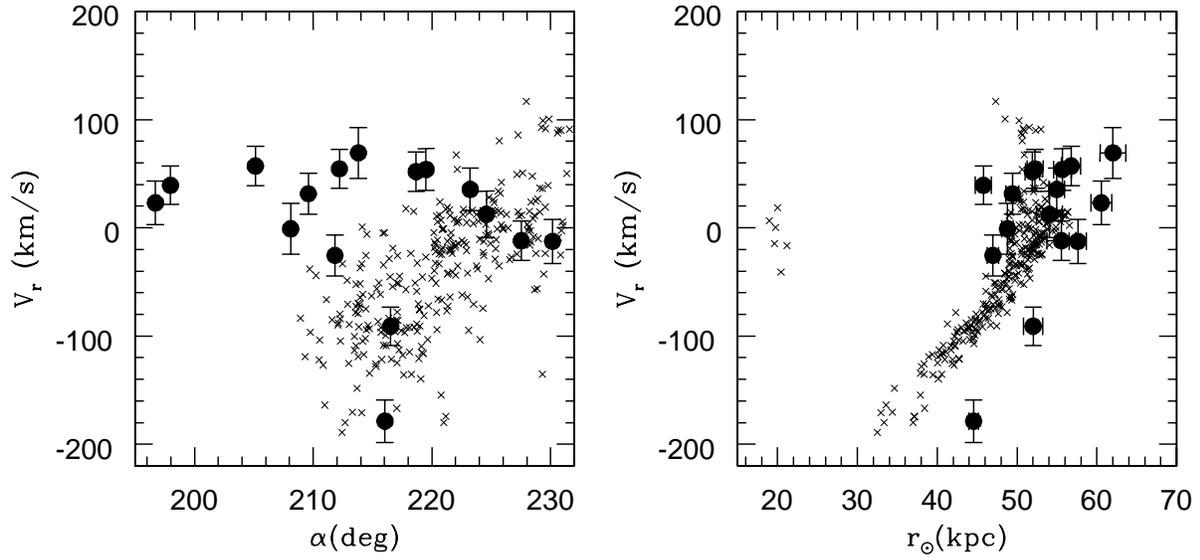}
\caption{The heliocentric radial velocitites of the RR Lyrae stars
(solid circles) are plotted as functions of right ascension and
distance from the Sun. The $\times$'s are the Sgr stream particles in
the model of \citet{mar04}.  The left plot only includes model particles
with distances $30<r_\sun <65$ kpc.
\label{fig-martinez}}
\end{figure}

\paragraph{Radial Velocity Distributions.}
In Figure~\ref{fig-martinez}, the radial velocities of the RR Lyrae
variables (solid circles) are plotted against right ascension and
distance from the Sun.  Also plotted are the particles in the
preferred model of Mart{\'\i}nez-Delgado et al.  This model predicts a
significantly lower mean velocity ($-39$ km/s) than our observations
and more significantly, large gradients in velocity with $\alpha$ and
with $r_\sun$ that are not observed.  Although the two RR Lyrae
variables that are outliers in Figure~\ref{fig-rvhist} have velocities
that overlap with many of the model particles, they lie at
significantly greater distances than the model.
 
\begin{figure}
\plotone{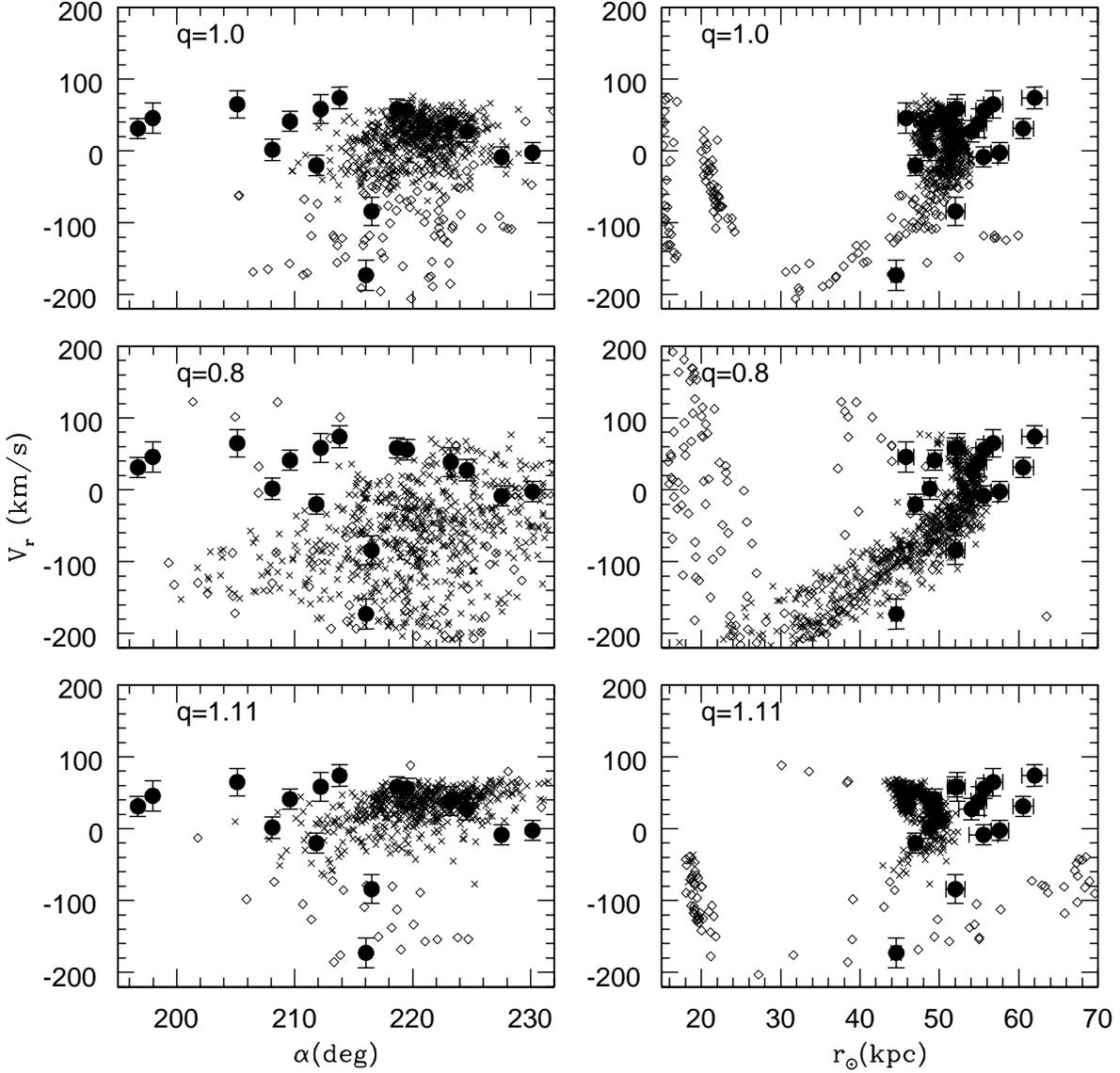}
\caption{Same plots as Figure~\ref{fig-martinez} but with the three
models of \citet{hel04} that assume different flattenings of the dark
matter halo of the Milky Way.  The $\times$'s are particles that
were liberated from the galaxy during the last 3.5 Gyrs. Open diamonds are
particles that became unbound between 3.5 and 7 Gyrs ago. The left
plots only include particles with distances $30<r_\sun <65$ kpc. Note
that the spherical (q=1.0) and prolate (q=1.11) halo models reproduce
the narrow velocity and distance distributions better than the oblate
(q=0.8) one.  However, all the models fail to match the distribution
in $\alpha$.
\label{fig-helmi}}
\end{figure}

Figure~\ref{fig-helmi} shows similar plots for Helmi's models.  The
oblate model with q=0.80 presents the same inconsistencies as the
Mart{\'\i}nez-Delgado et al.'s one: different mean velocity and
gradients in velocity with both $\alpha$ and $r_\sun$.  In contrast,
both the spherical and the prolate halo models have some appealing
similarities. Most of their particles became unbound from the main
body of Sgr during the past 3.5 Gyrs ($\times$'s).
Consequently, their radial velocity distributions are quite
narrow. The spherical model has $\langle V_h \rangle = 17$ km/s and
$\sigma = 28$ km/s, while the prolate halo has $\langle V_h \rangle =
28$ km/s and $\sigma = 26$ km/s. These numbers are very similar to our
measurements ($\langle V_h \rangle = 33$ km/s; $\sigma = 25$
km/s). Also, the two radial velocity outliers can be explained as
stars that became unbound between 3.5 and 7 Gyrs ago (open diamonds).  The
ratio of the oldest particles to the more recent parts of the stream 
in Figure~\ref{fig-helmi} (0.15 and
0.08 for q=1.0 and q=1.11 respectively) is consistent with our
observation that only 2 out of the 16 RR Lyrae stars have velocities
$<-80$ km/s.

The oblate models are clearly inferior to the spherical and prolate
models in predicting the observed narrow widths of the stream in
$r_\sun$ and in radial velocity.  However, all of the models fail to
predict the long extent of the stream in $\alpha$, and for this reason,
we are hesitant to conclude that the dark matter halo must be either
spherical or prolate.  Until more models are constructed with the goal
to fit the data presented here, it is unclear whether these mismatches
constitute fatal flaws or ones that can be removed with better choices
for the model parameters.  While the results of the above comparisons
are inconclusive, they do illustrate that the coupling of radial
velocity measurements with the precise distance information afforded
by RR Lyrae variables is a powerful diagnostic for testing models of
the stream\footnote{During the revision process of the present paper,
\citet{hel04b} and \citet{joh04} have arrived
at very different conclusions about the shape of the dark
matter halo from comparisons of the M giant streams with model
calculations.}.

\bigskip

Each of the above models predicts the presence of stream stars in the
same part of the sky but at closer distances ($15-25$ kpc). Because the
number density of these stars is much less than in the clump at 50
kpc, they will not constitute so obvious over-densities with respect
to the background of halo stars. The QUEST survey has detected
numerous bright RR Lyrae stars in the region, and some of them appear
to be clustered in weak over-densities \citep{vz04}. Spectroscopy of
these relatively bright RR Lyrae stars is underway with the goal of
seeing if they constitute moving groups.  Their possible
association with the Sgr stream will be examined.

\section{SUMMARY}

Our spectroscopic observations have yielded radial
velocity and metallicity measurements for a subsample of the RR Lyrae
variables that were discovered by the QUEST survey in the leading arm
of the Sgr tidal stream.  These data and the positional information
provided by the whole sample of QUEST RR Lyrae variables in the stream
provide powerful diagnostics to test numerical simulations of the Sgr
tidal stream.  

While recent models of the destruction of the Sgr galaxy reproduce the
general properties of the Sgr tidal streams, they fail to reproduce
the details revealed by the RR Lyrae variables.  This is particularly
true of models that assume a oblate flattening of the dark matter
halo.  Models that assume spherical and prolate dark matter halos
provide better fits to the data, but ones that are still marred by
some inconsistencies.  The most striking of these is the failure of
any model to reproduce is the observed span of the stream in right
ascension ($36\degr$ or $\sim 30$ kpc).

Our metallicity measurements show that the RR Lyrae stars in this part
of the stream belong to a metal-poor population with $\langle {\rm[Fe/H]}
\rangle = -1.77$.  This mean value is consistent with the age-metallicty
relation observed by \citet{lay00} in the central parts of the Sgr
galaxy.

\acknowledgments This work is based on observations collected with the
FORS2 instrument at VLT, Paranal Observatory, Chile (Project
69.B-0343A).  The data were obtained as part of an ESO service mode
run.  This research is part of a joint project between Universidad de
Chile and Yale University, partially funded by the Fundaci\'on Andes.
Financial support was also provided by the National Science Foundation
under grant AST-0098428. 
C.G. acknowledges
partial support from the Spanish Ministry of Science and Technology
(Plan Nacional de Investigaci\'on Cient{\'\i}fica, Desarrollo e 
Investigaci\'on
Tecnol\'ogica, AYA2002-01939) and from the European Structural Funds.
We warmly thank Amina Helmi, David
Mat{\'\i}nez-Delgado and Mari\'angeles G\'omez-Flechoso for making
available to us their models of the Sgr tidal streams, and Andrew Layden
for his programs for measuring the pseudo-equivalent widths of the
spectral lines. We also thank the anonymous referee for helpful suggestions 
to the manuscript.

Facilities: \facility{VLT(FORS2)}

\clearpage

\clearpage

\clearpage

\clearpage

\clearpage

\clearpage

\clearpage

\clearpage

\clearpage

\clearpage

\clearpage
\begin{deluxetable}{lcccccccccrcccc}
\tabletypesize{\scriptsize}
\rotate
\tablecolumns{15}
\tablewidth{0pt}
\tablecaption{Observation Log and Individual Results}
\tablehead{
\colhead{Star} & \colhead{$\alpha$ (2000.0)} & \colhead{$\delta$ (2000.0)} &
\colhead{V} & \colhead{Date} & \colhead{HJD} & \colhead{Texp} &
\colhead{Seeing} & \colhead{Airmass} & \colhead{Phase} & \colhead{$V_r$} &
\colhead{$\sigma_r$} & \colhead{W(K)} & \colhead{W(H)} & \colhead{[Fe/H]} \\
\colhead{} & \colhead{(h m s)} & \colhead{($\degr$ $\arcmin$ $\arcsec$)} &
\colhead{(mag)} & \colhead{} & \colhead{(+245000 d)} & \colhead{(s)} &
\colhead{($\arcsec$)} & \colhead{} & \colhead{} & \colhead{(km/s)} &
\colhead{(km/s)} & \colhead{\AA} & \colhead{\AA} & \colhead{} }
\startdata
244 &   13 06 48.49 &   -01 17 53.2 &   19.48 & 2002-06-03 &    2429.51027 &    2x600 & 1.69 &  1.11 & 0.600 &   62 & 16 & 4.16 & 4.15 & -1.67 \\
    &               &               &         & 2002-06-17 &    2443.55672 &    2x600 & 1.21 &  1.15 & 0.950 &   42 & 16 & \nodata & \nodata & \nodata \\
    &               &               &         & 2002-08-01 &    2488.49578 &    2x600 & 0.95 &  1.48 & 0.653 &   32 & 15 & 3.65 & 3.99 & -1.87 \\
252 &   13 11 52.35 &   -00 22 32.2 &   18.88 & 2002-06-03 &    2429.55508 &    2x540 & 1.36 &  1.10 & 0.125 &    1 & 15 & 2.53 & 5.52 & -1.95 \\
280 &   13 40 31.28 &   -00 41 11.5 &   19.40 & 2002-08-01 &    2488.51983 &    2x600 & 1.07 &  1.48 & 0.366 &   49 & 15 & 4.18 & 4.29 & -1.64 \\
294 &   13 52 25.11 &   -00 17 26.4 &   19.10 & 2002-08-08 &    2495.51402 &    2x540 & 1.54 &  1.54 & 0.769 &   24 & 18 & \nodata & \nodata & \nodata  \\
    &               &               &         & 2002-08-09 &    2496.50118 &    2x540 & 1.29 &  1.45 & 0.441 &    2 & 15 & 4.57 & 3.59 & -1.66 \\
304 &   13 58 24.02 &   -00 28 18.4 &   19.18 & 2002-07-14 &    2470.50192 &    2x540 & 1.05 &  1.13 & 0.594 &   30 & 15 & 4.72 & 3.80 & -1.57 \\
    &               &               &         & 2002-07-15 &    2471.50440 &    2x540 & 1.02 &  1.14 & 0.196 &   27 & 15 & 5.32 & 4.17 & -1.29 \\
308 &   14 07 20.91 &   -01 31 15.8 &   18.99 & 2002-07-14 &    2470.52379 &    2x540 & 0.93 &  1.15 & 0.689 &   -1 & 15 & 3.54 & 3.78 & -1.94 \\
    &               &               &         & 2002-08-10 &    2497.49844 &    2x540 & 1.25 &  1.36 & 0.379 &  -31 & 15 & 3.55 & 3.89 & -1.92 \\
312 &   14 08 49.81 &   -00 04 21.6 &   19.25 & 2002-07-14 &    2470.54993 &    2x600 & 0.93 &  1.24 & 0.060 &   10 & 15  & 1.42 & 7.89 & -1.81 \\
    &               &               &         & 2002-08-11 &    2498.51654 &    2x600 & 0.92 &  1.54 & 0.253 &   29 & 15 & 3.87 & 5.13 & -1.54 \\
321 &   14 15 13.55 &   -00 53 02.8 &   19.66 & 2002-08-01 &    2488.54836 &    2x600 & 1.40 &  1.51 & 0.290 &   27 & 16 & 4.36 & 4.77 & -1.46 \\
    &               &               &         & 2002-08-09 &    2496.52373 &    2x600 & 1.15 &  1.50 & 0.375 &   81 & 16 & 4.17 & 4.14 & -1.67 \\
333 &   14 24 10.68 &   -00 47 56.3 &   18.90 & 2002-07-15 &    2471.52481 &    2x480 & 1.25 &  1.14 & 0.776 & -141 & 16 & 3.78 & 3.61 & -1.90 \\
    &               &               &         & 2002-08-08 &    2495.53578 &    2x480 & 1.69 &  1.52 & 0.096 & -161 & 15 & 2.70 & 6.14 & -1.73 \\
337 &   14 26 05.67 &   -00 45 25.2 &   19.22 & 2002-07-08 &    2464.63663 &    2x600 & 0.77 &  1.65 & 0.661 &  -65 & 15 & 3.66 & 3.55 & -1.94 \\
    &               &               &         & 2002-07-14 &    2470.61572 &    2x600 & 0.75 &  1.60 & 0.056 & -113 & 15 & 1.85 & 7.14 & -1.83 \\
354 &   14 34 43.64 &   -01 13 08.8 &   19.20 & 2002-07-14 &    2470.57187 &    2x540 & 0.78 &  1.24 & 0.735 &   82 & 15 & 3.43 & 3.62 & -2.00 \\
    &               &               &         & 2002-08-10 &    2497.51996 &    2x540 & 1.11 &  1.38 & 0.215 &   29 & 15 & 3.69 & 4.50 & -1.75 \\
360 &   14 38 02.76 &   -01 28 22.9 &   19.36 & 2002-07-07 &    2463.61869 &    2x600 & 1.09 &  1.38 & 0.620 &   72 & 15 & 4.35 & 4.21 & -1.60 \\
    &               &               &         & 2002-08-09 &    2496.54597 &    2x600 & 0.94 &  1.54 & 0.106 &    8 & 15 & 1.72 & 7.12 & -1.90 \\
    &               &               &         & 2002-08-10 &    2497.54076 &    2x600 & 1.07 &  1.52 & 0.873 &   52 & 15 & \nodata & \nodata & \nodata \\
372 &   14 52 54.85 &   -01 20 48.3 &   19.35 & 2002-06-17 &    2443.70287 &    2x540 & 1.29 &  1.51 & 0.310 &   16 & 16 & 3.70 & 3.98 & -1.85 \\
    &               &               &         & 2002-08-04 &    2491.54559 &    2x540 & 1.15 &  1.35 & 0.045 &    1 & 15 & 2.13 & 6.59 & -1.85 \\
379 &   14 58 22.05 &   -01 02 02.6 &   19.40 & 2002-07-06 &    2462.59702 &    2x600 & 1.04 &  1.19 & 0.517 &   11 & 15 & 4.72 & 3.80 & -1.57 \\
    &               &               &         & 2002-08-11 &    2498.53914 &    1x600 & 0.92 &  1.41 & 0.163 &   13 & 18 & \nodata & \nodata & \nodata  \\
    &               &               &         & 2002-08-12 &    2499.50930 &    2x600 & 1.13 &  1.26 & 0.962 &  -29 & 15 & 3.28 & 5.60 & -1.64 \\
391 &   15 10 08.94 &   -00 58 29.2 &   19.48 & 2002-07-07 &    2463.66267 &    2x600 & 1.23 &  1.55 & 0.313 &  -22 & 15 & 3.38 & 5.45 & -1.64 \\
    &               &               &         & 2002-07-10 &    2466.56781 &    2x600 & 1.46 &  1.13 & 0.365 &  -33 & 15 & 3.69 & 5.01 & -1.64 \\
409 &   15 20 39.63 &   -00 00 09.4 &   19.39 & 2002-07-13 &    2469.60861 &    2x600 &  \nodata    &  1.26 & 0.794 & 4 & 15 & 2.49 & 3.63 & -2.29 \\
    &               &               &         & 2002-07-17 &    2473.60295 &    2x600 & 1.03 &  1.29 & 0.478 &   23 & 16 & 2.27 & 3.56 & -2.37 \\
\enddata
\label{tab-log}
\end{deluxetable}

\clearpage
\begin{deluxetable}{lrcclc}
\tablecolumns{6}
\tablewidth{0pt}
\tablecaption{Radial Velocity Standard Stars}
\tablehead{
\colhead{Star} & \colhead{V} & \colhead{Sp Type} &
\colhead{$V_h$} & \colhead{Date} & \colhead{Texp}\\
\colhead{} & \colhead{(mag)} & \colhead{} &
\colhead{(km/s)} & \colhead{} & \colhead{(s)} }
\startdata
Kopff 27  & 10.21 & A3V & +5.5  & 2002-06-03 & 15; 60 \\
Feige 56  & 11.10 & A0V & +30   & 2002-06-04 & 30; 90 \\
HD 97783  &  9.04 & G1V & +87.9 & 2002-06-04 & 2; 10 \\
HD 155967 &  7.42 & F6V & -15.8 & 2002-06-18 & 5  \\
\enddata
\label{tab-standards}
\end{deluxetable}

\clearpage
\begin{deluxetable}{lcrrcrcrcc}
\tablecolumns{10}
\tablewidth{0pt}
\tablecaption{Radial Velocities, Abundances and Distances 
of the RR Lyrae Stars}
\tablehead{
\colhead{Star} & \colhead{$N_{\rm fit}$} & \colhead{$V_\gamma$} &
\colhead{$\sigma$ (fit)} & \colhead{$\sigma_\gamma$} & \colhead{$V_{GSR}$} &
\colhead{$\Lambda_{\sun}$} & \colhead{$B_{\sun}$} & \colhead{[Fe/H]} &
\colhead{r$_\sun$} \\
\colhead{} & \colhead{} & \colhead{(km/s)} &
\colhead{(km/s)} & \colhead{(km/s)} & \colhead{(km/s)} & \colhead{($\degr$)} &
\colhead{($\degr$)} & \colhead{} & \colhead{(kpc)} }
\startdata
244  &  2  &  31  &  18     & 14 & -44 & 273.0 &  9.6 & -1.77 & 61 \\
252  &  1  &  46  & \nodata & 21 & -23 & 273.7 &  8.2 & -1.95 & 46 \\
280  &  1  &  65  & \nodata & 19 & 13  & 280.1 &  4.9 & -1.64 & 57 \\
294  &  2  &   2  &   9     & 15 & -41 & 282.4 &  3.1 & -1.66 & 49 \\
304  &  2  &  41  &  22     & 14 & 1   & 283.8 &  2.5 & -1.43 & 49 \\
308  &  2  & -20  &   3     & 14 & -57 & 286.3 &  2.2 & -1.93 & 47 \\
312  &  1  &  58  & \nodata & 20 & 26  & 285.9 &  0.8 & -1.68 & 52 \\
321  &  2  &  74  &  22     & 15 & 44  & 287.7 &  0.7 & -1.57 & 62 \\
333  &  1  & -173 & \nodata & 21 & -197& 289.5 & -0.5 & -1.82 & 45 \\
337  &  1  & -84  & \nodata & 20 & -106& 289.9 & -0.8 & -1.89 & 52 \\
354  &  2  &  58  &   5     & 14 & 40  & 292.0 & -1.5 & -1.88 & 52 \\
360  &  2  &  56  &   1     & 14 & 40  & 292.9 & -1.7 & -1.75 & 56 \\
372  &  1  &  39  & \nodata & 20 & 33  & 296.0 & -3.7 & -1.85 & 55 \\
379  &  2  &  27  &  23     & 15 & 26  & 297.0 & -4.6 & -1.61 & 54 \\
391  &  2  &  -9  &   9     & 14 & -2  & 299.6 & -6.1 & -1.64 & 56 \\
409  &  2  &  -3  &  27     & 24 & 14  & 301.4 & -8.3 & -2.33 & 58 \\
\enddata
\label{tab-rv}
\end{deluxetable}

\end{document}